\documentclass[
 reprint,
superscriptaddress,
 amsmath,amssymb,
prl,
floatfix,
]{revtex4-2}

\usepackage{graphicx}
\usepackage{dcolumn}
\usepackage{bm}
\usepackage[colorlinks,urlcolor=blue,citecolor=blue,linkcolor=blue]{hyperref}

\begin{document}

\preprint{topo}

\title{Topological phase transition in an all-optical exciton-polariton lattice}

\author{M.~Pieczarka}
 \email{maciej.pieczarka@pwr.edu.pl}
 \affiliation{ARC Centre of Excellence in Future Low-Energy Electronics Technologies and Nonlinear Physics Centre, Research School of Physics, The Australian National University, Canberra, ACT 2601, Australia}
 \affiliation{Department of Experimental Physics, Wroc\l{}aw University of Science and Technology, Wyb.~Wyspia\'nskiego 27, 50-370 Wroc\l{}aw, Poland}
 
\author{E.~Estrecho}%
 \affiliation{ARC Centre of Excellence in Future Low-Energy Electronics Technologies and Nonlinear Physics Centre, Research School of Physics, The Australian National University, Canberra, ACT 2601, Australia}

\author{S.~Ghosh}
\affiliation{Division of Physics and Applied Physics, School of Physical and Mathematical Sciences, Nanyang Technological University, Singapore 637371, Singapore}%

\author{M.~Wurdack}%
 \affiliation{ARC Centre of Excellence in Future Low-Energy Electronics Technologies and Nonlinear Physics Centre, Research School of Physics, The Australian National University, Canberra, ACT 2601, Australia}

\author{M.~Steger}
\thanks{current address: National Renewable Energy Laboratory, Golden, CO 80401, USA}
\affiliation{Department of Physics and Astronomy, University of Pittsburgh, Pittsburgh, PA 15260, USA}%
\author{D.~W.~Snoke}
\affiliation{Department of Physics and Astronomy, University of Pittsburgh, Pittsburgh, PA 15260, USA}%
\author{K.~West}
\affiliation{Department of Electrical Engineering, Princeton University, Princeton, NJ 08544, USA}%
\author{L.~N.~Pfeiffer}
\affiliation{Department of Electrical Engineering, Princeton University, Princeton, NJ 08544, USA}
\author{T.~C.~H.~Liew}
\affiliation{Division of Physics and Applied Physics, School of Physical and Mathematical Sciences, Nanyang Technological University, Singapore 637371, Singapore}
\author{A.~G.~Truscott}%
 \affiliation{Laser Physics Centre, Research School of Physics, The Australian National University, Canberra, ACT 2601, Australia}
 \author{E.~A.~Ostrovskaya}
 \email{elena.ostrovskaya@anu.edu.au}
 \affiliation{ARC Centre of Excellence in Future Low-Energy Electronics Technologies and Nonlinear Physics Centre, Research School of Physics, The Australian National University, Canberra, ACT 2601, Australia}

\maketitle

\noindent \textbf{Topological insulators \cite{Hasan2010} are a class of electronic materials exhibiting robust edge states immune to perturbations and disorder. This concept has been successfully adapted in photonics  \cite{Haldane2008,Khanikaev2016,Ozawa2019}, where topologically nontrivial waveguides \cite{Hafezi2013, Kruk2019, Rechtsman2013} and topological lasers \cite{Han2019, Ota2018, Bahari2017, Bandres2018,St-Jean2017} were developed. However, the exploration of topological properties in a given photonic system is limited to a fabricated sample, without the flexibility to reconfigure the structure {\em in-situ}. Here, we demonstrate an all-optical realization of the orbital Su-Schrieffer-Heeger (SSH) model \cite{Su1979} in a microcavity exciton-polariton system \cite{Deng2002,Kasprzak2006,Deng2010,Byrnes2014}, whereby a cavity photon is hybridized with an exciton in a GaAs quantum well. We induce a zigzag potential for exciton polaritons all-optically, by shaping the nonresonant laser excitation, and measure directly the eigenspectrum and topological edge states of a polariton lattice in a nonlinear regime of bosonic condensation. Furthermore, taking advantage of the tunability of the optically induced lattice we modify the intersite tunneling to realize a topological phase transition to a trivial state. Our results open the way to study topological phase transitions on-demand in fully reconfigurable hybrid photonic systems that do not require sophisticated sample engineering.}

Microcavity exciton polaritons (polaritons therein), hybrid quasiparticles resulting from strong coupling of excitons and photons in a semiconductor microcavity \cite{Deng2002,Kasprzak2006,Deng2010,Byrnes2014}, have emerged as a perfect platform for numerous applications in nonlinear and topological photonics \cite{Bardyn2016,Leykam2016,Ozawa2019, Klembt2018}. These interacting bosons combine a very low effective mass inherited from cavity photons with repulsive interactions inherited from excitons, allowing for bosonic condensation at elevated temperatures. 

Taking advantage of the photonic part of a polariton, one can modify the planar microcavity by various fabrication techniques and realize polariton trapping potentials, as well as a lattice of coupled traps \cite{Schneider2017,Whittaker2019,Klembt2018,St-Jean2017}. Additionally, the TE-TM polarized modes splitting in a planar cavity results in an effective spin-orbit interaction for polaritons enabling realizations of topological and flat-band systems \cite{Sala2015,Klembt2018,Whittaker2019,Alyatkin2020}. Nevertheless, this technological approach has a major practical drawback as once the sample is made, there is little or no room for modification of its properties. This limits the applications of polariton-based photonic topological devices, where active control is highly desirable \cite{Shalaev2019,Leykam2018,Kudyshev2019,Su2020}.

The solution to this issue lies in using the excitonic component of the polariton for engineering the trapping potential. Under nonresonant optical excitation above the semiconductor material bandgap, the pumping laser creates a high-energy excitonic reservoir \cite{Sanvitto2011}, which acts as a non-Hermitian potential that replenishes and repels polaritons due to exciton-polariton interactions \cite{Askitopoulos2013,Gao2015,Schneider2017}. The excitonic potential can therefore be shaped via spatially structured laser excitation.

\begin{figure*}[ht]
\centering
\includegraphics[width=\textwidth]{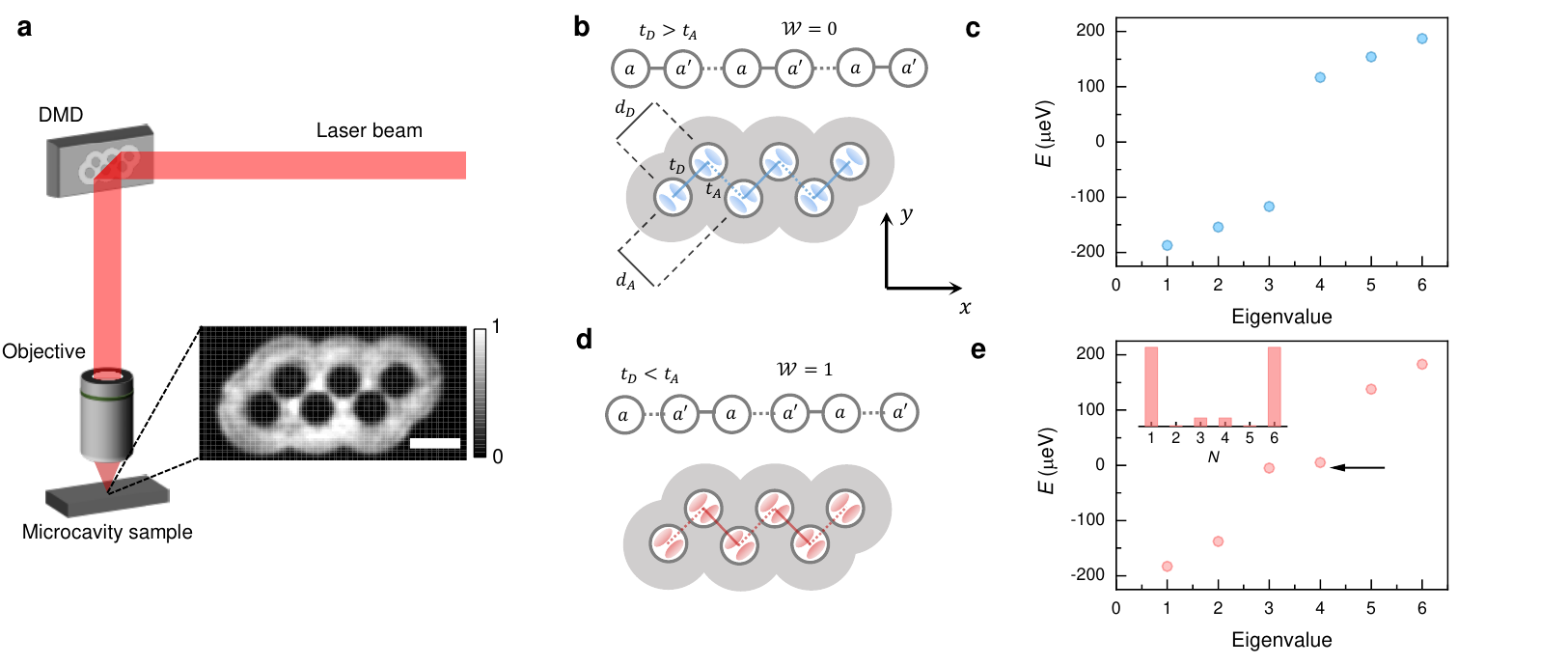}
\caption{\textbf{Realization of an orbital SSH Hamiltonian with a nonresonant optical excitation.} \textbf{a}, Simplified scheme of the experimental setup for creating exciton polaritons in an optically-induced trapping potential. Inset presents the spatial distribution of the laser pump reflected from the sample. Dark areas correspond to polariton traps with the trap diameter $D_{trap}\approx 5.9~\mu$m, arranged into a zigzag chain. White scale bar corresponds to 10 $\mu$m. \textbf{b,d}, Sketch of the orbital SSH model for (\textbf{b}) topologically trivial and (\textbf{d}) nontrivial cases, realized with the trapped $p$-modes of different orientations $A,D$. \textbf{c,e} Eigenenergies of a tight-binding Hamiltonian model corresponding to the $N=6$ chain in (\textbf{b,d}). Edge states within the gap are indicated with an arrow in panel (\textbf{e}). Inset shows the probability density distribution of the edge state.}
\label{fig:Fig1}
\end{figure*}

In this work, we employ a spatially structured laser beam, imaged via a microscope objective onto a planar microcavity sample with embedded GaAs quantum wells, see Fig.~\ref{fig:Fig1}(a). Spatial structuring is achieved by selectively reflecting the laser beam from a programmed digital micromirror device (DMD) (see Methods). The spatial distribution of the pump effectively creates a chain of coupled circular traps in the zigzag geometry, first proposed for the realization of low-dimensional topological systems in plasmonics and nanophotonics \cite{Poddubny2013,Sinev2015,Slobozhanyuk2015,Kruk2017,Trpathi2020}. Exciton polaritons trapped in this optically induced chain realize the orbital version of the Su-Schrieffer-Heeger (SSH) model, previously demonstrated in etched samples with coupled micropillars \cite{St-Jean2017,Whittaker2019,Harder2020}. 

The SSH chain \cite{Su1979} is the simplest realization of a topological insulator in one dimension (1D). Here, we focus on the SSH model realized with the $p$-modes of each circular trap, coupled as shown in Fig.~\ref{fig:Fig1}(b,d). In contrast to linear chains with alternating distances between the lattice sites \cite{Han2019, Ota2018, Pickup2020}, the orthogonality of the $p$-modes makes the SSH model valid for $d_A=d_D=d$, where $d_A$ and $d_D$ are the lattice constants in the antidiagonal ($A$) and diagonal ($D$) directions \cite{St-Jean2017}. This is because the tunneling amplitudes in these two directions, $t_A$ and $t_D$, are different due to the collinear or orthogonal alignment of the $p$-modes with the axis linking the consecutive traps. As a result, two configurations exist in finite length chains with an even number of lattice sites, with the staggering order being trivial $t_D>t_A$ for diagonal, $p_D$, and non-trivial $t_D<t_A$ for antidiagonal, $p_A$, mode orientations. The two different configurations (phases) are presented in Fig.~\ref{fig:Fig1}(b,d). They are characterized by phase winding numbers $\mathcal{W}=0$ for trivial and $\mathcal{W}=1$ for topological one (see Methods). The eigenenergies of a tight binding Hamiltonian model for finite sized chains investigated in this work are presented in Fig.~\ref{fig:Fig1}(c,e). The normal phase in Fig.~\ref{fig:Fig1}(c) is characterised by a spectrum with a trivial bandgap. The topologically nontrivial phase, presented in Fig.~\ref{fig:Fig1}(e),  differs from Fig.~\ref{fig:Fig1}(c), as two eigenvalues close to zero energy emerge in the bandgap. The corresponding eigenstates are strongly localized at the edges of the chain [see inset in Fig.~\ref{fig:Fig1}(e)] and are an indicator of a topological phase. In short chains, the spectra are discrete and the edge states can have energies slightly different from zero, depending on the ratio of the tunnelling amplitudes $t_D/t_A$. Nevertheless, the topological properties of the system are maintained.

\begin{figure*}[ht]
\centering
\includegraphics[width=\textwidth]{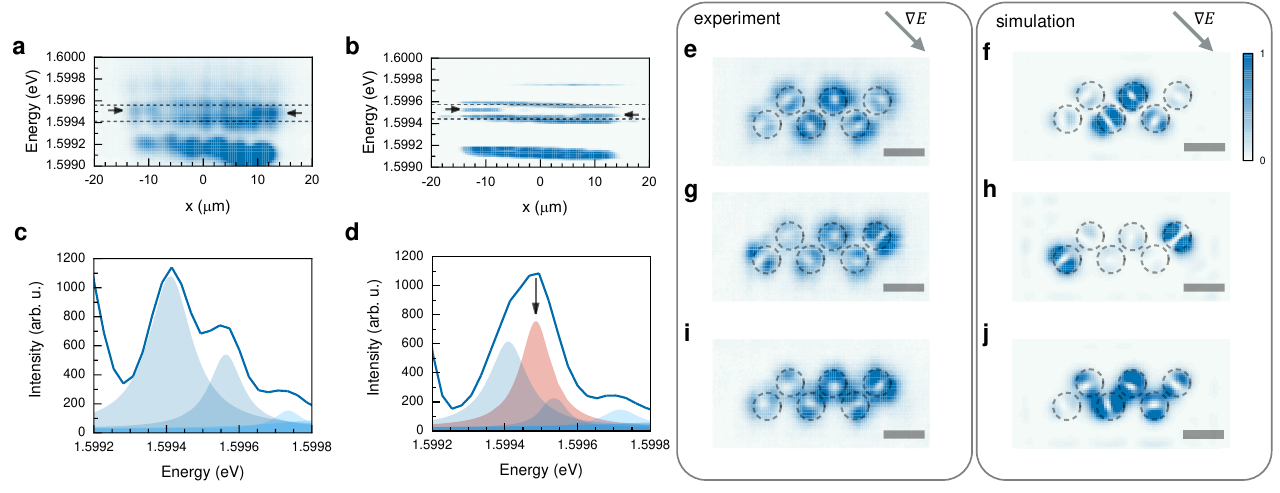}
\caption{\textbf{Position-resolved spectrum and density distribution of exciton polaritons in the SSH chain.} \textbf{a}, Experimentally measured spectrum along the chain with $N=6$ sites, integrated over the orthogonal direction. \textbf{b}, Corresponding spectrum obtained by numerical simulations of the mean-field open-dissipative model, Eq. (4) in Methods. \textbf{c,d}, Experimental spectrum measured at a position (\textbf{c}) in the centre of the chain, and (\textbf{d}) at the left edge of the chain. Shaded areas represent the result of fitting with Lorentzian lines. $p$-bands and $d$-band peaks are in semitransparent blue and the edge mode peak is coloured in semitransparent red. \textbf{e-i}, Experimental images of the exciton-polariton emission corresponding to the spatial density distributions taken at the energies of the (\textbf{e}) upper $p$-band, (\textbf{g}) edge states (middle of the bandgap) and (\textbf{i}) the lower $p$-band. \textbf{f-j}, Density distributions obtained by numerical simulations of the model Eq. (4) corresponding to the panels (\textbf{e-i}). The scale bar corresponds to 10 $\mu$m and the direction of the energy gradient in the sample is indicated with an arrow.}
\label{fig:Fig2}
\end{figure*}

We investigate the optically-induced SSH chain at pump powers slightly above the exciton-polariton condensation threshold. In this regime, the condensate occupies high-symmetry points of the momentum-space band structure (for details, see the data in Fig.~S1 of the Supplementary Information). The narrowing of the linewidth of the exciton-polariton emission above the condensation threshold allows us to resolve the ground and excited bands with the corresponding bandgaps. The low-energy $s$-band is highly populated due to efficient energy relaxation of polaritons towards the ground state \cite{Wouters2010}. We also observe nonzero occupation of excited states forming a higher-energy band. The corresponding position-resolved spectrum is presented in Fig.~\ref{fig:Fig2}(a). The intense signal from the $s$-band at low energies is clearly separated from the $p$-band by a bandgap. It indicates that the created potential confines the excited states and realizes the physics discussed in Fig.~\ref{fig:Fig1}. More importantly, the $p$-band is split into two sub-bands corresponding to the bonding (in-phase) and anti-bonding (out-of-phase) coupling between the lattice sites (in analogy to electron orbitals in molecules). The gap between the two $p$-bands is larger than the linewidth, which allows us to directly identify the localized states at the edges of the SSH chain. The energies of these states lie inside the $p$-band gap, as indicated by the arrows in Fig.~\ref{fig:Fig2}. The spectral cross-sections in the bulk (i.e., in the middle) and at the edge of the chain are presented in Figs.~\ref{fig:Fig2}(c,d), where the shaded areas present the result of fitting the spectrum with Lorentzian lines. Existence of the in-gap edge modes confirmed by these measurements is a signature of a topological phase of the SSH lattice \cite{Hasan2010, Whittaker2019, St-Jean2017, Harder2020}. We note that the occupation of all bands is not spatially homogeneous, as the sample is characterized with a intrinsic linear energy gradient due to the spatially varying thickness of the cavity (see Methods), oriented antidiagonally with respect to the $x$-axis in the presented data. This effect is captured in our simulations of the full open-dissipative mean-field model [see Methods, Eq. (4)], with the results presented in Fig.~\ref{fig:Fig2}(b), showing an excellent agreement with the experimental observations. We emphasize that the simulations are based on experimentally determined parameters of the sample (see Methods). The energy gradient does not change the topology of the system, as the polariton states at the individual lattice sites are hybridized, which is reflected in the opening of the bandgap (see also Fig.~S1). 

The spatial distribution of the exciton-polariton density in the chain is obtained by real-space spectral tomography, which enables selective real-space imaging of the polariton emission intensity at a given energy (see Methods). The experimental results are presented in Fig.~\ref{fig:Fig2}(e,g,i) together with the results of numerical modelling of Eq. (4) in Fig.~\ref{fig:Fig2}(f,h,j). The lower $p$-band shows a characteristic spatial distribution of a bonding state, where the $p$-modes from neighbouring traps overlap [Fig.~\ref{fig:Fig2}(i,j)]. The upper $p$-band shows an antibonding character with pronounced density dips between the traps indicative of the nodes in the probability density distribution [Fig.~\ref{fig:Fig2}(e,f)]. Finally, the edge states are composed of antidiagonal $p$-mode configuration forming the topologically nontrivial realization of the SSH model, Fig.~\ref{fig:Fig2}(g,h).

\begin{figure}
\centering
\includegraphics[width=\columnwidth]{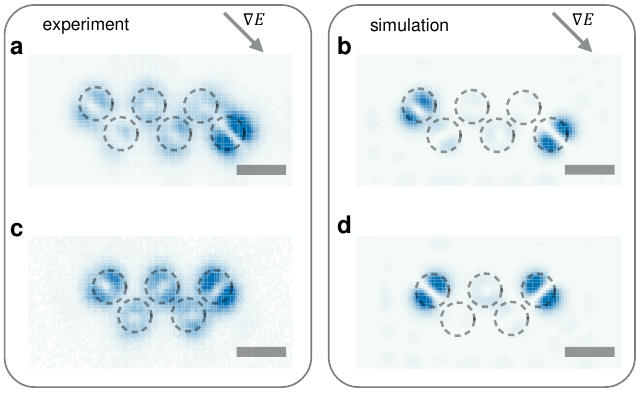}
\caption{\textbf{Topological edge states in different lattice realizations.} \textbf{a,c}, Experimental spatial density distributions of the edge state for (\textbf{a}) a chain of $N=6$ sites with a different orientation compared to the chain in Fig.~\ref{fig:Fig2} and (\textbf{c}) a chain of $N=5$ sites. \textbf{b,d},  Results of numerical simulations of the model Eq. (4) corresponding to the cases in (\textbf{a}) and (\textbf{c}). 
The scale bar corresponds to 10 $\mu$m and the direction of the energy gradient is indicated with an arrow.} 
\label{fig:Fig3}
\end{figure}

To test the robustness of the topologically protected edge states, we imprinted different chains at different positions on the sample with similar detunings. The edge state for the $N=6$ chain similar to that in Fig.~\ref{fig:Fig2}, but with a different orientation with respect to the energy gradient, is presented in Fig.~\ref{fig:Fig3}(a) with the corresponding numerical simulation shown in Fig.~\ref{fig:Fig3}(b). For this orientation, the diagonal $p$-mode configuration is the topologically nontrivial one. Similarly to the $N=6$ case, the orbital SSH model with an odd number of sites supports two edge states, however each edge state comes from a different $p$-mode configuration \cite{St-Jean2017,Whittaker2019,Harder2020}. This is clearly seen in our experimental data for a chain with $N=5$ sites, presented in Fig.~\ref{fig:Fig3}(c), in agreement with simulations in Fig.~\ref{fig:Fig3}(d). These results demonstrate the insensitivity of the all-optical realization of a topological SSH model to the orientation of the energy gradient and local disorder of the sample.

\begin{figure*}[ht]
\centering
\includegraphics[width=\textwidth]{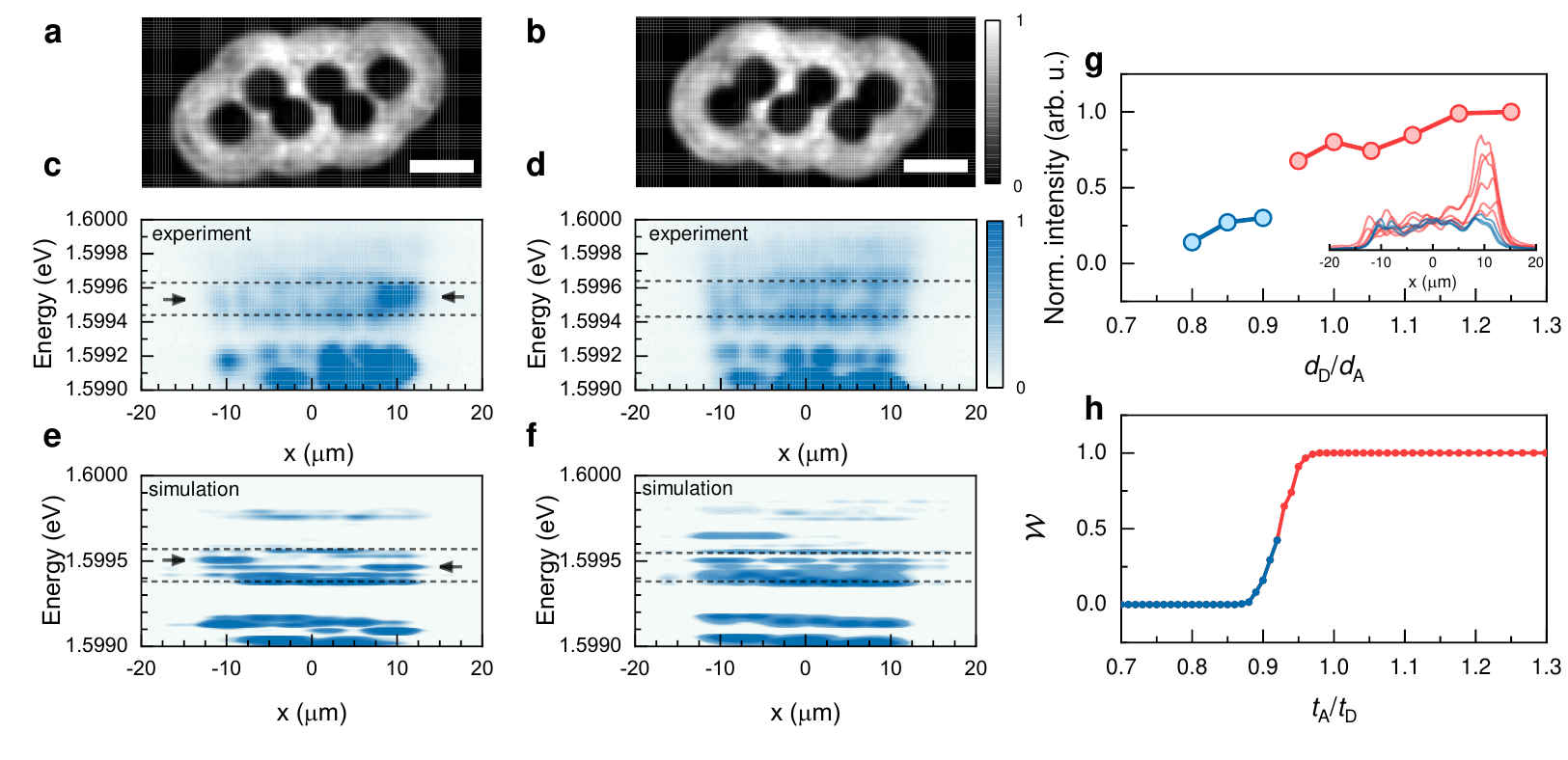}
\caption{\textbf{Measurement of the topological phase transition in SSH chains.} \textbf{a,b}, Spatial distributions of the excitation intensity reflected from the sample for the modified chains in the (\textbf{a}) topological and (\textbf{b}) the trivial phase. The scale bar corresponds to 10 $\mu$m.  \textbf{c,d}, Experimental spectra corresponding to the cases presented in (\textbf{a,b}). \textbf{e,f}, Numerically computed spectra corresponding to the experimental measurements in (\textbf{c,d}). \textbf{g}, Intensity of the exciton-polariton edge state at the right end of the modified chains as a function of $d_D/d_A$. Inset shows the intensity distributions measured in the middle of the $p$-band gap. \textbf{h}, Winding number $\mathcal{W}$ calculated from the tight-binding Hamiltonian, including the energy gradient and disorder [see Methods, Eqs. (1) and (3)], as a function of $t_A/t_D$. The trivial phase is coloured in blue and the topological phase is coloured in red in (\textbf{g,h}). }
\label{fig:Fig4}
\end{figure*}

To demonstrate an optically-driven topological phase transition in our system, we modified the topology of the $N=6$ chain by changing the ratio between the tunneling amplitudes from $t_A/t_D>1$ to $t_A/t_D<1$ in the $N=6$ chain. The coupling between the nearest-neighbour sites depends on the potential barrier amplitude as well as on the distance between the traps \cite{Galbiati2012}. Therefore, we increase the coupling $t_A$ or $t_D$ by reducing the trap separations in $A$ or $D$ directions, while keeping all other parameters constant. In this way, we modified the SSH Hamiltonian for $p_A$ configuration. Fig.~\ref{fig:Fig4}(a,b) presents the intensity distribution of the laser excitation reflected from the sample for the chains of modified dimerisations. The chain maintaining the topological phase is presented in Fig.~\ref{fig:Fig4}(a) and the chain in the trivial phase is shown in Fig.~\ref{fig:Fig4}(b). In both cases, the lattice constant in one of the directions was reduced by 20\%. The measured real-space spectra for these geometries are shown in Fig.~\ref{fig:Fig4}(c,d). One observes an increase of the topological and the trivial bandgaps between the $p$-bands in comparison to the unmodified chain: $\Delta E_{\text{topo}} = 174 \pm 10~\mu\text{eV}$ for $d_D/d_A = 0.8$ and $\Delta E_{\text{triv}} = 218 \pm 8~\mu\text{eV}$ for $d_D/d_A = 1.25$ in comparison to $\Delta E_{\text{topo}} = 151 \pm 9~\mu\text{eV}$ for $d_D/d_A = 1.0$, from Fig.~\ref{fig:Fig2}. Growth of this value for the modified chains is a direct manifestation of the control over the tunneling amplitudes, as the bandgap in the SSH model is proportional to $|t_D - t_A|$ (full set of values is presented in Supplementary Information). The geometrical modification leading to the change of the tunneling amplitudes influences the $s$-band as well, which is now clearly split, and signatures of edge modes are visible in Fig.~\ref{fig:Fig4}(c). The asymmetry of the $p$-band gap values is caused by the energy gradient, which influences the sensitivity of $t_A$ and $t_D$ to the trap separation parameter. Additional input to the tunneling amplitudes comes from the intrinsic TE-TM splitting and weak birefringence of the sample \cite{Bieganska2020, Su2020}. 

The cavity gradient influences the occupation of the topological edge states in the measured spectrum in Fig.~\ref{fig:Fig4}(c), with one of the edge states dominating the spectrum. On the other hand, the spectrum in Fig.~\ref{fig:Fig4}(d) shows no signatures of the edge states, as expected for a topological phase transition to a trivial configuration with $\mathcal{W}=0$.

The flexibility of our all-optical potential allows us to tune the lattice geometry parameters continuously. Thus, we performed a series of measurements to pinpoint the topological phase transition in the chain. Direct measurement of the winding number $\mathcal{W}$ is challenging in our experimental configuration \cite{St-Jean2020}, hence we measure the intensity distribution at the mid-gap and observe the appearance of the edge states as a signature of the topological phase, see Fig.~\ref{fig:Fig4}(g). The transition to a trivial state occurs around $d_D/d_A \approx 0.9-0.95$, where we observe an abrupt change in the intensity at the edge of the chain [there is a clear asymmetry in the edge modes' occupations, see inset in Fig.~\ref{fig:Fig4}(g)]. We reproduce this observation by calculating the winding number in the tight-binding Hamiltonian, including the potential gradient and random disorder (see Methods), which softens the transition threshold and moves it away from the point $t_A/t_D=1$, Fig.~\ref{fig:Fig4}(h), as observed in the experiment.

To summarize, we have demonstrated an all-optically driven topological phase transition in a fully reconfigurable optically-induced orbital SSH lattice created in an open-dissipative exciton-polariton system. The transition is controlled by fine-tuning the strength of tunneling between the lattice sites. We emphasize that implementing this kind of control would typically require fabrication of many different samples with different implementations of lattice geometries. Moreover, we have demonstrated the robust topologically protected edge states in a regime, where exciton polaritons are condensed and nonlinear (density-dependent) effects could begin to play a role \cite{Pernet2021}. Combined with the experimental control of the gain and loss (linewidth) of the trapped polariton condensates \cite{Gao2015} our system represents an attractive, flexible platform for further studies of topological effects in a nonlinear and non-Hermitian hybrid photonic system.

\section*{Methods}

\noindent \textbf{Sample and experiment} The sample used in the experiment is a high-quality GaAs-based microcavity with a long cavity photon lifetime exceeding 100 ps \cite{Pieczarka2020}. The cavity of the length $3\lambda/2$ is enclosed between distributed Bragg reflectors with 32 (top) and 40 (bottom) Al$_{0.2}$Ga$_{0.8}$As/AlAs layer pairs. The active region is made of 12 GaAs/AlAs quantum wells (QWs) of 7 nm nominal thickness positioned in three groups at the maxima of the confined photon field. The measured Rabi spliting is about $\hbar\Omega = 15.9 \pm 0.1$~meV. The experiments are done at slightly varying positions on the sample that correspond to the same photon-exciton detuning $\Delta = -0.43$~meV. All results are obtained with a microcavity kept in a continuous flow helium cryostat, ensuring the sample temperature of 7-8K.

The nonresonant optical excitation was provided by a continuous-wave (CW) Ti: Sapphire laser (M Squared SolsTiS), tuned to the cavity reflectivity minimum above the QW bandgap. The Gaussian laser beam was transformed and focused to a top-hat distribution by a shaping lens (Eksma Optics GTH-5-250-4) and imaged onto the digital micromirror device (DMD). The shape of the lattice potential was encoded on the DMD which reflected the laser selectively and then imaged onto the sample via set of lenses and a microscope objective \cite{Gao2015}. Photoluminescence from the sample was collected with the same objective in the reflection geometry and imaged with a set of confocal lenses onto the slit of the spectrometer (Princeton Instruments IsoPlane 320), equipped with a 2D sensitive CCD (Andor iXon Ultra 888). The imaging lens in front of the spectrometer was mounted on a motorized stage, enabling the spectral tomography of the emission. The tomography was done by collecting a set of spectral images, scanning the full real-space emission by moving the image with respect to the entrance slit.

\noindent \textbf{Theory.} The SSH model is described by a tight-binding Hamiltonian:
\begin{equation}
    \hat{H}_{SSH} = \sum_i \left( t_D\hat{a}^\dagger_i \hat{a}'_i + t_A \hat{a}^\dagger_{i+1}\hat{a}'_i + h.c. \right) 
\end{equation}
where $\hat{a}_i$ and $\hat{a}'_i$ are the annihilation operators in the $i^\text{th}$ unit cell of a lattice. The eigenfunctions of the Hamiltonian $\hat{H}_{SSH}$ are given by $(\pm e^{-i\varphi(k)}, 1)$, with $k$ being the wave vector. The winding number $\mathcal{W}$ is  defined by,
\begin{equation}
 \mathcal{W} =  \frac{1}{2\pi} \int_\text{BZ} dk, \frac{\partial \varphi(k)}{\partial k}
\end{equation}
corresponding to the Zak phase $\mathcal{Z} = \pi \mathcal{W}$, which is a 1D lattice equivalent of the geometric Berry phase. In an ideal SSH model, $\mathcal{W}=1$ for $t_D<t_A$ and $\mathcal{W}=0$ for $t_D>t_A$. In our system, the lattice is perturbed by a potential gradient and  disorder. To include these effects, we consider a Hamiltonian given by: 
\begin{equation}
  \hat{H} = \hat{H}_{SSH}+ \sum_i \left( V_{2i-1} \hat{a}^\dagger_i\hat{a}_i + V_{2i}\hat{a}'^\dagger_i\hat{a}'_i \right).
\end{equation}
The potential energy is given by $V_i = v_i + v_0( i-N/2)$, where $v_0$ is a constant, $v_i$ is a random value representing disorder in the system, and $N$ is the number of unit cells in the lattice. For our simulations $v_0 = 0.2/N$, $v_i$ is randomly distributed within the interval $[\pm 0.05]$, and energies are expressed in the unit of $\text{max}(t_A,t_D)$. 

Within the mean-field approximation, exciton polaritons in optically induced potentials can be described by a driven-dissipative equation for the polariton wavefunction, $\psi(x,y,t)$:
\begin{eqnarray}
    i\hbar \frac{\partial \psi}{\partial t}= \left[-\lambda\frac{\hbar^2}{2m}\nabla^2 + iP + V(x,y) + \alpha|\psi|^2  \right] \psi
    \label{Eq:GP}
\end{eqnarray}
where $\lambda = (1-i\lambda_0)$, $\lambda_0$ is a phenomenological parameter describing the energy relaxation~\cite{Read2009}, and $m$ is the effective mass of the polaritons. The net gain $P(x,y) = P_0(x,y)-\gamma$ is given by the difference between the pump strength (gain) $P_0(x,y)$ and the loss $\gamma$ (linewidth). The parameter $\alpha = (\alpha_R-i\alpha_I)$ represents the polariton-polariton interaction $\alpha_R$ and nonlinear decay $\alpha_I$. $V(x,y)$ accounts for a real part of the potential (the SSH lattice) induced by the repulsive interactions of the polaritons with an optically injected excitonic reservoir, as well as for the energy gradient in the cavity. For the numerical simulations, we consider $\alpha_{R} = \alpha_{I}$ and a transformation $\psi \to \psi/\sqrt{\alpha_{R}}$ to obtain an equation describing the exciton polaritons: $i\hbar \partial\psi/\partial t = [-(\lambda\hbar^2)/(2m)\nabla^2 + iP(x,y) + V(x,y) + (1-1i)|\psi|^2] \psi$. The loss and effective mass parameters are obtained from the experimental data in~\cite{Pieczarka2020}. Considering the long lifetime of exciton polaritons in our GaAs-based sample, we use a linewidth $\gamma = 5.5~\mu \text{eV}$ corresponding to a lifetime $120$ps. We take $m=7.6\times10^{-5}m_e$, where $m_e$ is the mass of an electron. Other parameters were chosen as $\lambda_0 = 0.05~ \text{meV}$ and $\max(P_0) = 10.5~\mu\text{eV}$, to fit the phenomenology of the present experiment.

\subsection{Author contribution statement}
E.A.O. and A.G.T. supervised and guided this research. M.P. initiated the research project. M.P. and E.E. designed and performed the experiment. S.G. and T.C.H.L. performed theoretical analysis and numerical simulations. M.P. analysed and interpreted the data with the input from E.E., M.W., A.G.T, and E.A.O. M.S. and D.W.S. designed and characterised the sample, which was grown by  K.W. and L.N.P. M.P. and E.A.O. wrote the manuscript with the input from all authors.

\end{document}